\documentclass[aps,prl,tightenlines,twocolumn,showpacs,preprintnumbers,amsmath,amssymb]{revtex4}

\usepackage{graphicx}
\usepackage{dcolumn}
\usepackage{bm}
\usepackage{epsfig}

\begin{document}

\title{Planar SFS Josephson Junctions Made by Focused Ion Beam Etching.}


\author{V.M. Krasnov}
\author{ O. Ericsson}
\author{ S. Intiso}
\author{ P. Delsing}

\affiliation{Department of Microtechnology and Nanoscience,
Chalmers University of Technology, SE-41296 Gothenburg, Sweden}

\author{ V.A. Oboznov}
\author{ A.S. Prokofiev}
\author{ V.V. Ryazanov}

\affiliation{Institute of Solid State Physics, Russian Academy of
Sciences, 142432 Chernogolovka, Russia}

\begin{abstract}

Superconductor-Ferromagnet-Superconductor (S-F-S) Josephson
junctions were fabricated by making a narrow cut through a S-F
double layer using direct writing by Focused Ion Beam (FIB). Due
to a high resolution (spot size smaller than 10 nm) of FIB,
junctions with a small separation between superconducting
electrodes ($\leq$ 30 nm) can be made. Such a short distance is
sufficient for achieving a considerable proximity coupling through
a diluted CuNi ferromagnet. We have successfully fabricated and
studied S-F-S (Nb-CuNi-Nb) and S-S'-S (Nb-Nb/CuNi-Nb) junctions.
Junctions exhibit clear Fraunhofer modulation of the critical
current as a function of magnetic field, indicating good
uniformity of the cut. By changing the depth of the cut, junctions
with the $I_c R_n$ product ranging from 0.5 mV to $\sim 1\mu $V
were fabricated.

\end{abstract}


\maketitle

\section{Introduction}

Superconducting circuits, containing hybrid Superconductor -
Ferromagnet (S-F) structures are of considerable interest for
possible applications in cryoelectronics. Exchange interaction in
a ferromagnet results in splitting of electron spectra for spin up
and - down orientations. Therefore, Cooper pairs, consisting of
electrons with opposite spin acquire the momentum $±Q=E_{ex}/v_F$,
where $E_{ex}$ is the exchange energy in the ferromagnet and $v_F$
is the Fermi velocity, when they penetrate into the ferromagnet
via the proximity effect at the S-F interface. This leads to an
oscillating superconducting order parameter $\psi \propto
cos(2Qx)$, which in turn could lead to a sign reversal of the
order parameter in the S-F-S structure and formation of the
$\pi$-junction, exhibiting a spontaneous $\pi$-shift in the phase
difference or the negative Josephson coupling, when the thickness
of the F-layer is close to an odd integer number of half the
oscillation period \cite{1}. Structures including arrays of
conventional 0- and $\pi$-junctions can operate as "phase
batteries" which can minimize interactions with the dephasing
environment and could allow realization of a "quiet" phase qubit-
the basic element of a quantum computer \cite{2} or can be used
for building complimentary Josephson digital devices \cite{3} and
novel modification of Rapid Single Flux Quantum logic \cite{4}.

Such S-F-S $\pi$-junctions were fabricated recently
\cite{RyazPRL,Kontos} and existence of a spontaneous $\pi$-shift
in a triangular SFS-junction array \cite{RyazPRB} and one-junction
interferometers \cite{Frolov} was also demonstrated. However,
there are still considerable technical difficulties, which have to
be overcome before such junctions could be used in practical
devices. One of the difficulties is associated with a short
coherence length in a ferromagnet:

\begin{equation}
\xi_F^{dirty}=\sqrt{\frac{\hbar D}{E_{ex}}},
\end{equation}

where $D$ is a diffusion coefficient of electrons in a dirty
ferromagnet. For common ferromagnets with $E_{ex} \sim $ 1000K,
$\xi_F$ is in the range of a nanometer, as observed, eg., for pure
Ni \cite{Blum,Villegas}. For the sake of reproducibility, the
roughness of such a ferromagnetic film has to be kept at the
atomic level. This impose very strong demands on fabrication
techniques. To avoid this difficulty, diluted ferromagnetic alloys
with a reduced $E_{ex}$ can be employed. Approximately a ten-fold
increase of the ferromagnetic coherence length in diluted
CuNi\cite{RyazPRL} and PdNi\cite{Kontos} alloys was achieved,
resulting in a successful fabrication of S-F-S $\pi$-junctions
using conventional thin film deposition techniques. Another
technical difficulty is caused by a necessity to fabricate both 0-
and $\pi$- junctions at the same chip in one run, required for
practical quantum device.

In this paper we report a novel fabrication technique for planar
S-F-S junctions by Focused Ion Beam (FIB) etching. FIB allows
fabrication of nano-scale structures and has a great flexibility,
which is important for fabrication of more complicated circuits
containing 0 and $\pi$-junction arrays. Previously it was
demonstrated that FIB can be used for fabrication of
proximity-coupled Superconductor-Normal metal-Superconductor
(Nb-Cu-Nb) junctions\cite{Moseley} and SQUID's\cite{Burnell}. Here
we fabricate Superconductor-Ferromagnet-Superconductor Josephson
junctions by making a narrow FIB cut through a S-F double-layer.
Due to a high resolution (spot size smaller than 10 nm) of FIB,
junctions with a small separation between superconducting
electrodes ($\leq$ 30 nm) can be easily made. Such short distance
is sufficient to achieve a considerable proximity coupling through
a diluted CuNi ferromagnet. We have successfully fabricated and
studied S-F-S (Nb-CuNi-Nb) and S-S'-S (Nb- Nb/CuNi-Nb) junctions.
Junctions exhibit clear Fraunhofer modulation of the critical
current as a function of magnetic field, indicating good
uniformity of the cut. Changing the depth of the cut, junctions
with the $I_c R_n$ product ranging from 0.5 mV to $\sim 1 \mu V$
were fabricated.

In comparison to fabrication techniques for sandwich (overlap)
type S-F-S junctions reported so far\cite{RyazPRL,Kontos}, our
technique for fabrication of planar S-F-S junction using FIB is
much simpler and considerably more flexible. This technique can be
used for fabrication of 0 or $\pi$-junctions and $0-\pi$-junction
SQUIDs in one run by changing the depth and the width of the cut.

\section{Sample fabrication}

Cu$_{0.47}$Ni$_{0.53}$/Nb bilayers were deposited on oxidized Si
substrates by RF and DC magnetron sputtering, respectively. The
Cu$_{0.47}$Ni$_{0.53}$ alloy had a Curie temperature $\sim 60 K$
and the coherence length $\xi_F \sim$ 2-7 nm, Eq. (1). The bilayer
was patterned by optical lithography and Ar ion milling to define
4-5 $\mu$m wide electrodes and contact pads. The sample was then
transferred to a standard FIB (FEI Inc. FIB-200) for the junction
fabrication. The Ga ion beam had a dwell time of 0.3 $\mu$s, a
beam spot overlap of 30$\%$, and an acceleration voltage of 30 kV
through the process. The top panel in Fig.1 shows a sketch of the
fabrication procedure. After focusing and correcting the
astigmatism, a single cut through the top Nb layer was made at the
beam current of 1 pA. The width of the cut is determined by the
FIB spot size ($<$ 10nm). Finally, at the beam current of 10 pA,
rectangular patterns at the edges of electrodes were etched to
avoid electrical shorts from the material resputtered during Ar
ion milling.

The bottom panel in Fig.1 shows a secondary electron image of the
junction. The width of the cut at the top of Nb layer is $\sim$ 30
nm. We believe that walls of the cut are not completely vertical
and the actual width of the cut at the CuNi layer is narrower
(down to the spot size of $\sim$ 10 nm). The depth of the cut was
altered by changing the etching time. The etching rates of Nb and
CuNi were estimated to be $\sim 10^{-9} m^3/C$ by etching $2\times
2 \mu m^2$ square at the 10 pA beam current and using the end
point detection. However, we experienced that the etching rate
becomes somewhat smaller for deep cuts with the width-to-depth
aspect ratio more than two. Most probably, this is caused by
resputtering of material inside the trench, which is also partly
responsible for the walls of the cut being non-vertical. To change
the aspect ratio of the cut required for a complete etching
through the Nb layer, and thus to change the effect of
resputtering, we have studied junctions made from two types of
Nb/CuNi double layers with either 70 or 25 nm thick Nb layer. The
thickness of the CuNi layer was always 50 nm. Measurements were
done in a four probe configuration in a He$^4$ cryostat or a
He$^3$/He$^4$ dilution refrigerator.

\begin{figure}
\begin{minipage}{0.48\textwidth}
\epsfxsize=0.9\hsize \leftline{ \epsfbox{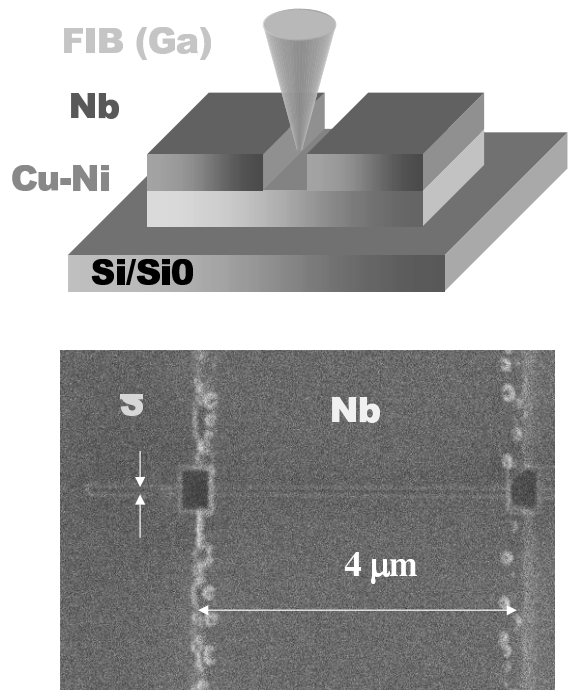}}
\caption{Top panel: A sketch of the fabrication procedure and the
sample geometry. Bottom panel: a secondary electron image of the
junction. } \label{Fig.1}
\end{minipage}
\end{figure}

\section{Results and Discussion}

Fig.2 shows a resistive transition of a Nb(70nm)/ CuNi(50nm)
junction. The inset shows the overall transition. Here the
resistance $\sim 100 \Omega$ is predominantly the resistance of
the bilayered Nb/CuNi bridge. It is seen that upon cooling down,
the resistance of the bridge first drops abruptly at $T_c = 7.5
K$, followed by a relatively broad transition which ends at $\sim
5-6 K$. Unlike the transition at $T_c$, the width of the second
transition changes from sample to sample and probably represents a
spontaneous flux-flow resistance of the Nb/CuNi
bridge\cite{RyazJETPL} caused by stray magnetic fields in the
ferromagnetic layer, which depend on a magnetic domain structure
of a particular sample. The resistive transition of the junction
itself is almost invisible in the inset of Fig.2 and is shown in
detail in the main panel of Fig.2. It is seen that the resistance
$R_n$ of the junction is much smaller than the resistance of the
bridge. For junctions with a measurable critical current, $R_n$
ranges from $\sim 0.2$ to $\sim 1.0 \Omega$, which is comparable
with the resistance of $\sim 0.1-0.2 \Omega$ for our CuNi alloy
with the resistivity $\rho \simeq 60 \mu\Omega cm$, the thickness
(the depth of the current path) $\sim 20-50 nm$, the length $\sim
20-30 nm$ and the width $\sim 4-5 \mu m$.

\begin{figure}
\begin{minipage}{0.48\textwidth}
\epsfxsize=0.9\hsize \leftline{ \epsfbox{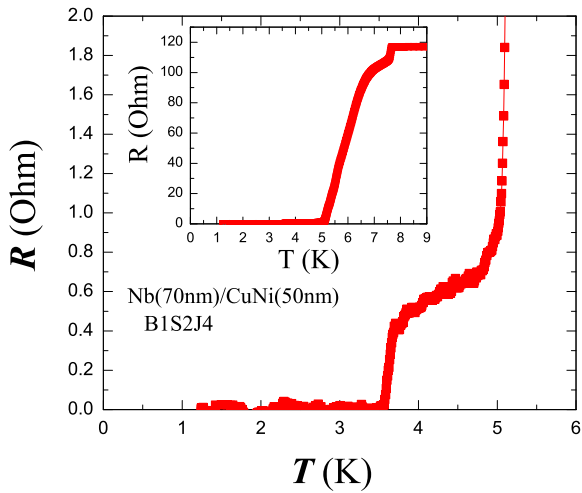}} \caption{A
typical resistive transition of a Nb(70nm)/CuNi(50nm) junction }
\label{Fig.2}
\end{minipage}
\end{figure}

Fig. 3 shows current voltage characteristics (IVC's) of another
Nb(70nm)/CuNi(50nm) junction at different base temperatures. Inset
shows IVC's at elevated temperatures. It is seen that the onset of
the critical current, $I_c$, is sharp and well defined even at
elevated temperatures. At lower temperatures IVC's exhibit a
hysteresis, so that the retrapping current, $I_r$, at which the
junction switch from the resistive to the superconducting state is
smaller than the critical current, $I_c$, see the IVC at $T=1.2K$
in Fig.3. With increasing temperature the hysteresis (the
difference between $J_c$ and $J_r$) vanishes as demonstrated in
Fig.4 for two Nb(25nm)/CuNi(50nm) junctions. A similar hysteresis
was also reported for Nb/Cu planar junctions and was attributed to
the self-heating phenomenon \cite{Burnell}. Indeed, self-heating
could be considerable for S-S'-S junctions with large $I_c$.
However, we observed a similar hystersis even in S-F-S type
junctions with the same geometry but with a three order of
magnitude smaller $I_c$ and proportionally smaller dissipation
power at the retrapping current, $P_r$. For example,  the junction
B2S1aJ1 with a large critical current $I_c = 1.58$mA and the
dissipation power $P_r=37.7$nW exhibits the hysteresis $I_c/I_r
=2.7$ at $T=30$mK; the junction B2S1aJ6 with an intermediate $I_c
= 175 \mu$A and $P_r=1$nW has $I_c/I_r =2.1$ at $T=30$mK, see
Fig.4; while the junction B2S1aJ5 Nb(25nm)/CuNi(50nm) with a small
$I_c=34.5 \mu$A and  $P_r=73.5$pW has $I_c/I_r =1.4$ at $T=30$mK,
i.e., the hysteresis has decreased by less than a factor two
despite the decrease of the dissipation power $P_r$ by 513 times.
All of those three junctions were fabricated on the same chip, had
the same geometry, except for the depth of the cut, and,
inevitably, similar thermal conductances, and were measured under
the same conditions. Therefore, it is unlikely that the self
heating alone could explain the observed hysteresis in IVC's. In
general, the hysteresis in IVC's at low temperatures is quite a
common phenomenon in superconductor-normal metal-superconductor
weak links and can also be attributed to non-equilibrium
phenomena\cite{Song,Tinkham} or frequency dependent
damping\cite{Kautz}. Recently a hysteresis in planar
superconductor-two dimensional electon gas-superconductor
junctions was shown to be dominated by a considerable stray
capacitance of electodes\cite{Thilo} .

\begin{figure}
\begin{minipage}{0.48\textwidth}
\epsfxsize=0.9\hsize \leftline{ \epsfbox{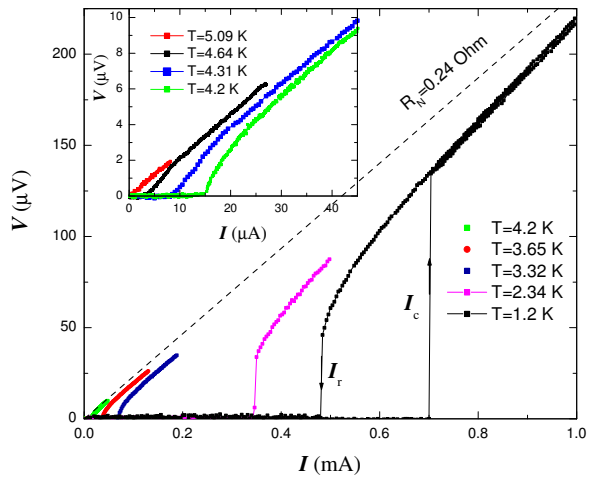}}
\caption{Current voltage characteristics at different temperatures
for a Nb(70nm)/CuNi(50nm) junction.} \label{Fig.3}
\end{minipage}
\end{figure}

In Fig.4, temperature dependencies of the linear critical current
density $J_c$ (A/m) are shown for Nb(70nm)/CuNi(50nm) and
Nb(25nm)/CuNi(50nm) junctions. It is seen that the critical
current density can be varied within three orders of magnitude by
changing the etching time, i.e. the depth of the cut. For example,
if we take the junction B1S1J4 as a reference point (etching time
80 sec. for a 6 $\mu$m long line), an increase of etching time by
12.5$\%$ (90 sec.) for the junction B1S1J3 leads to a two fold
decrease of $J_c (T=1.2K)$. However, further increase of the
etching time by $50 \%$ (120 sec) for the junction B1S1J2 leads to
a 70-fold drop of $J_c (T=1.2K)$. Apparently, such a dramatic drop
in $J_c$, which is associated with a small change of the etching
time (depth) and a negligible change in the junction resistance,
occurs at the threshold of a complete etching through the Nb
layer. Therefore, junctions B1S1J4 and B1S1J3 are of S-S'-S
(Nb-Nb/CuNi-Nb) type, where the weak link S' consists of a thin
underetched Nb layer with suppressed superconducting properties
due to proximity effect with the underlying CuNi alloy, while the
junction B1S1J2 is of S-F-S (Nb-CuNi-Nb) type. Taking into account
the difference in etching times and critical currents between
samples B1S1J4 and B1S1J3, and assuming that in B1S1J3 not more
than $ \xi_s \sim 10 n$m of Nb is left, we can estimate the
etching rate to be $\sim 0.36$ nm/sec per $\mu$m of the cut at the
FIB current of 1 pA, which is consistent with the measured etching
rate of $\sim 10^{-9} m^3/C$, for the width of the cut $\sim 30$
nm, see Fig.1. Similar estimations for B1S1J2 indicate that for
this junction we have etched not more than 5 nm into the CuNi
alloy.

The critical current continues to decrease rapidly with increasing
the depth of the cut into the CuNi. Here one has to compare the
depth of the cut into CuNi with the coherence length $\xi_F \sim
2-7 nm$. For deeper cuts effective length of the ferromagnetic
link in such planar S-F-S junctions becomes longer, causing the
rapid decrease of $J_c$.

\begin{figure}
\begin{minipage}{0.48\textwidth}
\epsfxsize=0.9\hsize \leftline{ \epsfbox{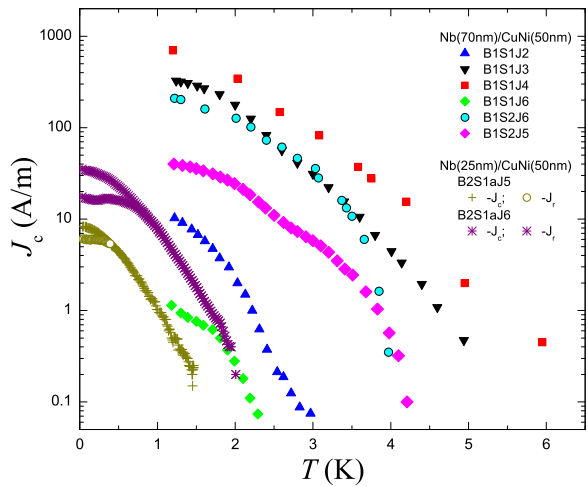}}
\caption{Temperature dependence of the linear critical current
density for junctions with different depth and width of the cut
and different thickness of the Nb electrode.} \label{Fig.4}
\end{minipage}
\end{figure}

In contrast to the critical current, the normal resistance of the
junction does not exhibit any peculiarity at the threshold of
etching through Nb. For example, the normal resistance increased
by merely $30\%$ from junction B1S1J3 ($R_n \simeq 0.26 \Omega$)
to junction B1S1J2 ($R_n \simeq 0.33 \Omega$), which is consistent
with estimation of the remaining thickness of the film after
cutting. For junctions with a smallest measurable critical current
density, $J_c \sim 0.1$A/m, $R_n$ increases to $\sim 1.0 \Omega$,
i.e., by approximately a factor of five in comparison with the
S-S'-S junction B1S1J4, while the critical current decreased more
than a thousand times. Since the critical current is much more
sensitive to the depth of the cut than the normal resistance, the
$I_c R_n$ product decreases with increasing the depth of the cut.
For junctions shown in Fig. 4, the $I_c R_n$ ranges from $\sim$
0.5 mV to $\sim 1 \mu$V at the lowest temperature, with the $I_c
R_n$ product at the threshold of cutting through the Nb layer
being $\sim 10-20 \mu V$.

From Fig.4 it is seen that the critical current has a strong
temperature dependence (note the logarithmic scale in $J_c$ axis)
with a positive curvature of $J_c(T)$ close to $T_c$. Such
behavior is typical for long Superconductor-Normal
metal-Superconductor junctions, in which the length of the normal
metal exceeds the coherence length, $\xi_N$ \cite{Golub}. This is
probably the case for our S-S'-S junctions, in which S' can behave
as a normal metal due to a strong proximity effect between a thin
Nb layer and the underlying CuNi film. However, we also observed a
strong $J_c (T)$ dependence for nominally S-F-S junctions, which
was unexpected, since for $E_{ex} \simeq 60 K$ for our CuNi alloy,
the coherence length $\xi_F$ has a negligible temperature
dependence in the corresponding temperature range, $\pi k T \ll
E_{ex}$. We could suggest that the strong $J_c(T)$ dependence can
be due to several reasons: first, if the walls of the cut are not
perfectly vertical, the current is flowing from regions with
gradually vanishing thickness of Nb, and therefore continuously
varying $T_c$ \cite{RyazJETPL}. It can also be caused by the
implantation of Ga into Nb and CuNi alloy, by the damage of
surface layers of both Nb and CuNi and by resputtering of material
inside the cut. All this might lead to local suppression of $T_c$
of Nb and formation of a thin non-magnetic layer at the surface of
CuNi alloy. The depth of implantation of Ga into Cu was estimated
to be $\sim$ 10 nm \cite{Moseley}. The resputtering can be reduced
by decreasing the thickness of Nb and, therefore, the aspect ratio
of the cut. As seen from Fig. 4, decreasing of the Nb thickness
from 70 to 25 nm does not introduce major changes in the $J_c(T)$
behavior, except for a reduced $T_c$, which indicates that
resputtering is probably not the major reason for the observed
strong $J_c(T)$ dependence.

Fig.5 shows a typical dependence of the critical current versus
magnetic field, $I_c(H)$, for a S-F-S type Nb(25nm)/CuNi(50nm)
junction B2S1aJ6 at $T=30mK$. Magnetic field was applied
perpendicular to the film, as required by the planar structure of
the junction, see Fig.1. It is seen that the $I_c(H)$ exhibits
clear Fraunhofer oscillations, indicating a good homogeneity of
the critical current, and thus homogeneity of the depth and the
width of the cut along the length of the junction. The periodicity
of Fraunhofer oscillations is about an order of magnitude smaller
than $\Delta H = \Phi_0/L\Lambda$, where $\Phi_0$ is the flux
quantum, $L$ is the length of the junction, and $\Lambda \simeq
2\lambda + d$- the effective magnetic width, where $\lambda$ is
the London penetration depth and $d$ is the width of the FIB cut.
The discrepancy is due to the fact that in our case magnetic field
is applied not parallel, but perpendicular to the superconducting
film. In this configuration flux focusing at the edges of the
superconducting film takes place and the effective magnetic field,
experienced by the junction, is $H_{eff}=H/(1-D)$, where $D$ is
the effective demagnetization factor of the superconducting
film\cite{KLR}. For a thin film in parallel magnetic field
$D_{\parallel}\simeq 0$ and flux focusing effect is negligible.
However in a perpendicular field $D_{\perp}\simeq 1$
\cite{Landau}, leading to a considerably smaller periodicity of
Fraunhofer modulation $\Delta H = \Phi_0(1-D_{\perp})/L\Lambda$.

\begin{figure}
\begin{minipage}{0.48\textwidth}
\epsfxsize=0.9\hsize \leftline{ \epsfbox{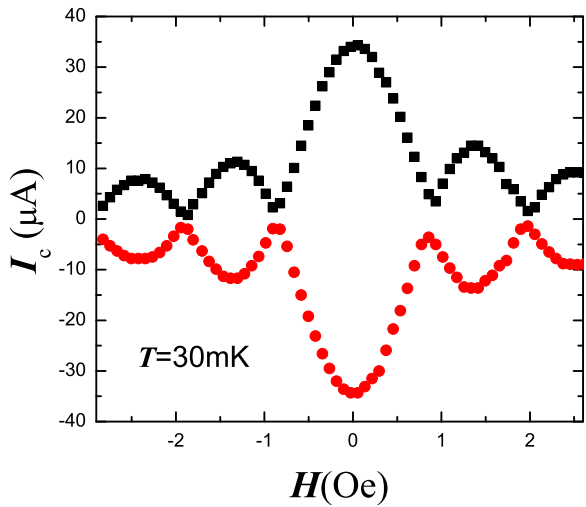}}
\caption{A Fraunhofer pattern $I_c(H)$ for a Nb(25nm)/CuNi(50nm)
junction.} \label{Fig.5}
\end{minipage}
\end{figure}

\section {Conclusions}
In conclusion, planar superconductor-ferromagnet-superconductor
Josephson junctions were fabricated by direct writhing with
Focused Ion Beam. Junctions exhibit clear Fraunhofer modulation of
the critical current as a function of magnetic field, indicating
good uniformity of the FIB cut. By changing the depth of the FIB
cut we could fabricate both variable thickness SS'S bridges with
$I_c R_n$ up to $0.5 mV$, when the cut is stopped within the
Nb-layer, and SFS junctions with $I_cR_n$ down to $1\mu V$, when
the Nb-layer is completely cut through. The cut through Nb layer
is accompanied by a dramatic drop of the critical current density
and only marginal change of the junction resistance. Finally we
argue that flexibility of FIB allows simultaneous fabrication of
"0-" and "$\pi-$" junctions in one run, by adjusting the depth of
the cut and the width of the junction. Such circuits would be
required for future applications of SFS junctions in novel
cryoelectronic devices.


\section{Acknowledgments}
The work was supported by INTAS-2001-0809 and Program of Russian
Academy of Sciences. We are grateful to T.Bauch for assistance in
experiment.

\end{document}